\documentclass[preprint,preprintnumbers]{revtex4}

\usepackage{graphicx}

\usepackage{amssymb}

\usepackage{mathrsfs}

\begin{document}

\noindent

\title{Two body interactions induces the  axion-phason field   in  Weyl semimetals }

\author{D. Schmeltzer}

\affiliation{Physics Department, City College of the City University of New York,  
New York, New York 10031, USA}

 \begin{abstract}
Following our results that the two-body interaction can induce a space and time dependent topological axion term $ \frac{e^2}{2\pi \hbar}\theta(z,t)(\vec{E}\cdot\vec{B})$, we show that  by applying this theory to a Weyl semimetal with two nodes in a magnetic field  a one-dimensional sliding charge density wave (CDW) in agreement with the recent experimental finding of J. Gooth et al. [Nature, https://doi.org/10.1038/s41586-019-1630-4 (2019)]. 
 Wee show  that the theory is equivalent to a time and space dependent of the topological angle which represents the phason. 
\end{abstract}

\maketitle

\textbf{I. INTRODUCTION}

Few years ago we showed that in topological insulator there is no  need to use dimensional projection from from $ 4+1$ dimension          to $ 3+1$ dimension \cite{Schmeltzer,Avadh} to derive the second Chern  number, instead we can use a two body  CDW interaction which has two solution.   A gap is open in the electronic spectrum  when the  $2k_{F} $ electronic  momentum   is commensurate  with the two body  interaction resulting    a topological term with a constant $\theta$  $ H=\int\,d^{3}x\theta(\frac{e^2}{2\pi \hbar})(\vec{E}\cdot\vec{B})$.   
  For  $\theta=\pm \pi$ the system is time reversal invariant  . The two body system allows an additional solution. The second solution corresponds to an incommensurate case  with an oder parameter   $\Delta =|\Delta|e^{i\alpha(\vec{r},t)}$ where $\Delta= 0$ and $|\Delta|\neq 0$. For this case the fluctuation of the phason $\alpha(\vec{r,t})$ generates the axion term  $\delta H=\int\,d^3x\theta(\vec{x},t)(\frac{e^2}{2\pi \hbar})(\vec{E}\cdot\vec{B})$.
A C.D.W.  theory \cite{Lee} has been apply by \cite{Goth} et al.
Other explanation based   on the Edelstein effect has  been given  \cite{Edelstein}.

We aim to present a microscopic theory for the case that the electric field and the magnetic field are parallel. For this case the low energy model is equivalent to a one dimensional Weyl Hamiltonian  which can be Bosonized. Including the two body interaction  we recover
the axion-phason model  for the  Weyl case.
The two $Weyl$ bands are
 linearized  around each Weyl node.  $2Q_{3}$ is the distance between the Weyl nodes in the momentum space.   The Hamiltonian  with the two nodes is given by: (for more 
than two nodes we consider that two nodes are aligned with the magnetic field)
\begin{eqnarray}
&&H^{R}=\int\,d^{3}x\Psi^{\dagger}_{R}(\vec{x})\Big[\hbar v \vec{\sigma}_{\perp}\cdot(\vec{k}_{\perp}-e\vec{A}_{\perp})+\hbar v\sigma_{3}(k_{3}-Q_{3})\Big]\Psi_{R}(\vec{x})
\nonumber\\&&H^{L}=\int\,d^{3}x\Psi^{\dagger}_{L}(\vec{x})\Big[-\hbar v \vec{\sigma}_{\perp}\cdot(\vec{k}_{\perp}-e\vec{_A}_{\perp})-\hbar v\sigma_{3}(k_{3}+Q_{3}\Big]\Psi_{L}(\vec{x})\nonumber\\&&
\end{eqnarray}
\noindent
Following \cite{Landsteiner}  the two component  spinor     can  be written as a  Dirac spinor :.
\begin{eqnarray}
&&\bar\Psi(\vec{x},t)=\Psi^{\dagger}(\vec{x},t)\gamma_{0}\nonumber\\&&
\vec{\gamma}=\vec{\sigma}\times i\tau_{2},\gamma_{0}=I\times\tau_{1}, \gamma_{5}=-I\times\tau_{3}\nonumber\\&&
\end{eqnarray}
 The Dirac Weyl  Hamiltonian takes the form:
\begin{eqnarray}
&&H^{Weyl}=\int \,d^3 x\bar\Psi(\vec{x},t)\Big[\hbar v\vec{\gamma}_{\perp}\cdot(\vec{k}_{\perp}-e\vec{A}_{\perp})+\hbar v\gamma_{3}(k_{3}-\gamma_{5}Q_{3})-\gamma_{0}\mu\Big]\Psi(\vec{x},t)\nonumber\\&&
\end{eqnarray}
The momentum $Q_{3}$ plays the role of the axial field   when $Q_{3} $ is  by space and time dependent field $Q_{3}(x,t)$ \cite{Landsteiner}
The basic property of the Weyl semimetals is the chiral anomaly which with the non-conserving the axial current .
	 The  axial anomaly is  obtained  from linear diverging  triangle diagram \cite{Nair} which consist from  two Electro-Magnetic  fields and the  third   axial field $Q_{3} $\cite{Landsteiner,Fujikawa} .In the presence of attractive two body  interactions    $-U_{eff.}n_{R}n_{L}$ 
		 we replace the interaction with a Hubbard Stratonovici  field   $  \Delta_{Real}(\vec{x},t)\bar\Psi(\vec{x},t)\Psi(\vec{x},t)+i\Delta_{Imag.}(\vec{x},t)\bar\Psi(\vec{x},t) \gamma_{5}\Psi(\vec{x},t)$
The Hamiltonian becomes:
\begin{eqnarray}
&&H^{Weyl}=\int \,d^3 x\bar\Psi(\vec{x},t)\Big[\hbar v\vec{\gamma}_{\perp}\cdot(i\partial_{x_{\perp}}-e\vec{A}_{\perp})+\hbar v\gamma_{3}(-i\partial_{z}-\gamma_{5}Q_{3}) +i\Delta_{Imag.}(\vec{x},t\gamma_{5}+M(\vec{x},t)-\gamma_{0}\mu\Big]\Psi(\vec{x},t)\nonumber\\&&
M(\vec{x},t)=\Delta_{Real}(\vec{x},t);\Delta_{Imag.}(\vec{x},t)\nonumber\\&&
\end{eqnarray}
We compute the triangle diagram in the presence of the mass term $M(\vec{x},t)$   and chiral field $ i\Delta_{Imag.}(\vec{x},t)\gamma_{5}$. The contribution to the effective action from fields $ \Delta_{Imag.}(\vec{x},t) \vec{A}_{\perp}$ becomes:
\begin{eqnarray}
&&S^{(3)}_{eff.}=\int\,d^{3}x_{1}\int\,dt_{1}\int\,d^{3}x_{2}\int\,dt_{2}\int\,d^{3}x_{3}\int\,dt_{3}\nonumber\\&&Tr\Big[G(\vec{x_{1}},t_{1};\vec{x_{2}},t_{2})\gamma_{5}\Delta_{Imag.}(\vec{x_{2}},t_{2})G(\vec{x_{2}},t_{2};\vec{x_{3}},t_{3})\vec{\gamma}_{perp}
\cdot\vec{A}_{\perp}(\vec{x_{3}},t_{3})\vec{\gamma}_{perp}
\cdot\vec{A}_{\perp}(\vec{x_{1}},t_{1})\Big]\nonumber\\&&
\end{eqnarray}
Performing the computation using  the Green's function $G(\vec{x},t;\vec{x'},t')$ we obtain a linear diverging result \cite{Nair}.
We regularize the effective  action ,replacing $\Delta_{Imag.}(\vec{x},t)\rightarrow +\Delta_{Imag.}(\vec{x},t)+\partial_{z}\beta(z,t)$.
We obtain:
\begin{eqnarray}
&&S^{(3))}_{eff.}=\int\,dt\int\, d^3x\Big[\frac{e^2}{h}\beta(z,t)(\vec{E}\cdot\vec{B})\Big]\nonumber\\&&
\end{eqnarray}
The action depends on the angle  between the electric field  and  the magnetic field which is a  function of the interaction -phason.The   phason  gives  the
  sliding CDW phase is the origin to  the axion field.
 In a recent paper \cite{Goth} it 
was shown that when E is paralel to  B,  from the  positive magneto conductivity we can deduce the 
 presence of the sliding phase phason-axion phase. In this paper we will not consider the pinning  phase by the impurities,for this reason  our theory will be applicable only  for $E>E_{treshold-pining}$

\vspace{0.2 in}




\vspace{0.2 in}

\textbf{II-The Weyl representation in the presence of  interactions }  

\vspace{0.2 in}

The magnetic field is in the ''Z"  direction.
For this case the longitidunal conductivity will be positive.
We will  consider the case where the $ Weyl$   two nodes are in the "Z'' direction .
 
For this case we can write a one dimensional model based on the $Weyl$
dispersion for the Landau level $n=0$ and $\sigma_{3}=-1$ for the right and left modes. 
We have the linear dispersion  $(k_{3}+Q_{3})$ (right moover) and $-(k_{3}-Q_{3})$(left moover) \cite{Ninomiya}. We mention that this description will hold  better  for $(TaSe_{4})_{2}I$ which  is a  quasi-one-dimensional  material.
Projecting the Hamiltonian  to the $n=0$ Landau level we obtain the one dimensional model :
\begin{eqnarray}
&&P_{n=0}H^{Weyl}P_{n=0}=H^{1d,Weyl}=\int\,dz\hbar v\Big[\tilde{
\Psi}^{\dagger}_{R,\sigma=\downarrow}(-i\partial_{z}+Q_{3})\tilde{
\Psi}_{R,\sigma=\downarrow}-\tilde{\Psi}^{\dagger}_{L,\sigma=\downarrow}(-i\partial_{z}-Q_{3})
\tilde{\Psi}_{L,\sigma=\downarrow}\Big]\nonumber\\&&
\end{eqnarray}
 We include the  two body attractive interaction projected to the $n=0$ Landau level $-U_{eff.}n_{R,sigma=\downarrow}n_{L,sigma=\downarrow}$
 Decoupling the interaction term we obtain,
\begin{eqnarray}
&&P_{n=0}H^{(int.)}P_{n=0}=\nonumber\\&&H^{(1d,int.)}=\int\,dz\Big[ \Delta(z)\tilde{\Psi}^{\dagger}_{R,\sigma=\downarrow}\tilde{\Psi}_{L,\sigma=\downarrow}+\tilde{\Psi}^{\dagger}_{L,\sigma=\downarrow}\tilde{\Psi}_{R,\sigma=\downarrow} +\frac{|\Delta(z)|^2}{4U_{eff.}}\Big]
\nonumber\\&&
\end{eqnarray}
We introduce the one dimensional  fields: 
\begin{eqnarray}
 &&\tilde{\Psi}_{R,\sigma=\downarrow}(z)=e^{i(k_{F}+Q_{3})z}C_{R}(z)
 ;\tilde{\Psi}_{L,\sigma=\downarrow}(z)=e^{-i(k_{F}+Q_{3})z}C_{L}(z)\nonumber\\&&
\end{eqnarray}
We use the Bosonization rule:
\begin{eqnarray}
&&C_{R}(z)=\sqrt{\frac{1}{2\pi a}}e^{i \sqrt{4\pi}\theta_{R}(z,t)}\nonumber\\&&C_{L}(z)=\sqrt{\frac{1}{2\pi a}}e^{-i \sqrt{4\pi}\theta_{L}(z,t)}\nonumber\\&&
\theta_{R}(z,t)+\theta_{L}(z,t)=2\theta(z,t);
\theta_{R}(z,t)-\theta_{L}(z,t)=2\varphi(z,t)=2\varphi(z,t)\nonumber\\&&
\end{eqnarray}
The charge density wave  order parameter is  given by,
\begin{eqnarray} 
&&<\Psi^{\dagger}_{L,\sigma=\downarrow}(z)\Psi_{R,\sigma=\downarrow}
(z)>=\Delta(z)=|\Delta|e^{i\alpha(z,t)}\nonumber\\&&
\end{eqnarray}
The Weyl Hamiltonian with the interaction term is:
\begin{eqnarray}
&&H^{(Weyl+int.)}=\int\,dz\hbar\Big[\frac{ v}{2}(\partial_{z}\varphi(z,t))^2+\frac{ v}{2}(\partial_{z}\theta(z,t))^2+\frac{|\Delta|}{\pi}\cos[\sqrt{4\pi}\theta(z,t)+\alpha(z,t)+(2Q_{3}+2k_{F})z]
\Big]\nonumber\\&&
\end{eqnarray} 
The Fermi momentum  $K_{F}$ is shifted by the nodal momentum $ Q_{3}$. 
The $ C.D.W.$ order parameter is given by  $\Delta=|\Delta|e^{i\alpha(z,t)}$   .In the comensurate case $\alpha(z,t)=qz$ with  $q
\approx-(2Q_{3}+2k_{F})$ and  $\Delta=\Delta^{*}$.In the incomensurate case $|\Delta|\neq 0$ and $\alpha(z,t)=
+\hat{\alpha}(z,t)+qz$
Under this conditions we expand $H^{(Weyl+int.)}$   around the minimum  which is given  by $\sqrt{4\pi}\theta(z,t)+\hat{\alpha(z,t)}\approx\pi$.
This gives  the phason action 
\begin{eqnarray}
&&H^{(Weyl+int.)}=\int\,dz\Big[\frac{K_{c}v_{c}}{2}(\partial_{z}(\partial_{z}\varphi(z,t))^2+\frac{v_{c}}{2K_{c}}(\partial_{z}\hat{\alpha}(z,t))^2.......\Big]
\hspace{0.2in}, K_{c }=\sqrt{4\pi}>1\nonumber
\\&&
\end{eqnarray} 
Since  the Luttinger parameter $K_{c}>1$ the model is in the gaples phase even in the presence of additional interaction.

\textbf{III- The sliding phase responce to an  Electric field} 

\vspace{0.2in} 

The sliding CDW phase proposed by \cite{Lee} has been observed by \cite{Goth}.

The  action in the presence of the electic field $E_{3}(z,t)=\partial_{z}a_{0}(z,t)-\partial_{t}a_{3}(z,t)>E_{treshold-pining}$\cite{Stone} takes   the form.
\begin{eqnarray}
&&S^{(Weyl+int.)}=\frac{1}{8\pi}\int\,dz\int\,dt\Big[\frac{1}{K_{c}}\Big(v_{c}(\partial_{z}\hat{\alpha}(z,t))^2-\frac{1}{v_{c}}(\partial_{t}\hat{\alpha}(z,t))^{2}\Big)\Big]\nonumber\\&&
S^{(E)}=\frac{e}{2\pi}\int\,dz\int\,dt\Big[-a_{0}\partial_{z}\hat{\alpha}(z,t)+a_{3}\partial_{t}\hat{\alpha}(z,t)\Big]\nonumber\\&&
\end{eqnarray}
We find the phason driven by the electric fiield  $E_{3}(z) $

\begin{eqnarray}
&&-\frac{1}{2}\partial_{z}\Big(\frac{v_{c}}{K_{c}}(\partial_{z}\hat{\alpha}(z)\Big)=eE_{3}(z)\nonumber\\&&
\end{eqnarray}

The conductivity due to the phason  will be ,

 $\sigma=\frac{e^2}{h}K_{c}$

\vspace{0.2in}

\textbf{III-The Chiral transformation}

\vspace{0.2in}

The efect of the interaction is equivalent to a shift of  the topological angle $Q_{3}$ . 
This can be seen from the following  chiral transformation:
\begin{eqnarray}
&&\Psi(\vec{x},z)'=e^{i\gamma_{5}\beta(z)}\Psi(\vec{x},z)\nonumber\\&&\gamma_{0}\Psi^{\dagger}(z)'=\gamma_{0}\Psi^{\dagger}(\vec{x},z)e^{i\gamma_{5}\beta(z)}\nonumber\\&&
\gamma_{0}=I\times\tau_{1}
\nonumber\\&&
\end{eqnarray}
Equation $(3)$ will be modified to $Q_{3}\rightarrow Q_{3}+\partial_{z}\beta(z)$. 
In $3+ 1$ dimension the triangle diagram is superficial linear divergent .This divergence can be eliminated by a  linear shift of substraction of the linear divergent  divergent triangle diagram \cite{Nair}. Performing the chiral transformation replaces $ Q_{3}$ to $Q_{3}+\partial_{z}\beta(z)$.This change of variables gives a finite change for  the action $3+ 1$ dimensions  .
\begin{eqnarray}
&&\delta S=\int\,d^{2}x\int\,dz\int\,dt\Big[(\frac{e^2}{h})\beta(z)(\vec{E}\cdot\vec{B})\Big]\nonumber\\&&
\end{eqnarray}
The phase $\beta(z)$i s fixed by the phason term.
 The one dimensional  interaction is modified  by the chiral transformation,  $\frac{|\Delta|}{\pi}\cos[\sqrt{4\pi}\theta(z,t)+\alpha(z,t)+2\beta(z)+(2Q_{3}+2k_{F})z]$

This equation  fixes the chiral phase $\beta(z)$

 $-\frac{1}{2}\alpha(z,t)=\beta(z,t)$

This  shows that the phason plays the  role of the axion.

\textbf{IV-Conclusions}

\vspace{0.2 in}
The axion-phason action  has been obtained for the  Weyl semimetal without relying on the phenomenological theory for charge density  wave \cite{Lee} or   
the asymptotic theory given by $\int\,d^{2}x\int\,dz\int\,dt \frac{e^2}{h}\beta(z,t)(\vec{E}\cdot\vec{B})$

\end{document}